\documentclass[conference]{IEEEtran}
\usepackage{cite}
\usepackage{graphicx}  
\usepackage{amssymb,times}

\usepackage{algorithm}
\usepackage{algorithmic}

\usepackage{multirow}
\usepackage{psfrag}

\usepackage[cmex10]{amsmath}
\usepackage{amssymb}
\usepackage{color}
\usepackage{amsthm}

\def\tr{\mathrm{tr}}

\newtheorem{theorem}{Theorem}

\newtheorem{assumption}{Assumption}

\usepackage[dvipsnames,svgnames]{xcolor}
\usepackage[textwidth=30mm]{todonotes}

\newcommand{\eye}[1]{\mathbf{I}_{#1}}
\renewcommand{\rm}[1]{\mathrm{#1}}

\newcommand{\be}{\begin{equation}}
\newcommand{\ee}{\end{equation}}
\newcommand{\ba}{\begin{array}}
\newcommand{\ea}{\end{array}}
\newcommand{\bea}{\begin{eqnarray}}
\newcommand{\eea}{\end{eqnarray}}
\newcommand{\herm}{^{{\rm{\scriptsize H}}}}

\newcommand{\tran}{^{\mbox{\scriptsize T}}}

\newcommand{\vbar}{\raisebox{.17ex}{\rule{.04em}{1.35ex}}}
\newcommand{\vbarind}{\raisebox{.01ex}{\rule{.04em}{1.1ex}}}
\newcommand{\R}{\ifmmode {\rm I}\hspace{-.2em}{\rm R} \else ${\rm I}\hspace{-.2em}{\rm R}$ \fi}
\newcommand{\T}{\ifmmode {\rm I}\hspace{-.2em}{\rm T} \else ${\rm I}\hspace{-.2em}{\rm T}$ \fi}
\newcommand{\N}{\ifmmode {\rm I}\hspace{-.2em}{\rm N} \else \mbox{${\rm I}\hspace{-.2em}{\rm N}$} \fi}
\newcommand{\B}{\ifmmode {\rm I}\hspace{-.2em}{\rm B} \else \mbox{${\rm I}\hspace{-.2em}{\rm B}$} \fi}
\newcommand{\Hil}{\ifmmode {\rm I}\hspace{-.2em}{\rm H} \else \mbox{${\rm I}\hspace{-.2em}{\rm H}$} \fi}
\newcommand{\C}{\ifmmode \hspace{.2em}\vbar\hspace{-.31em}{\rm C} \else \mbox{$\hspace{.2em}\vbar\hspace{-.31em}{\rm C}$} \fi}
\newcommand{\Cind}{\ifmmode \hspace{.2em}\vbarind\hspace{-.25em}{\rm C} \else \mbox{$\hspace{.2em}\vbarind\hspace{-.25em}{\rm C}$} \fi}
\newcommand{\Q}{\ifmmode \hspace{.2em}\vbar\hspace{-.31em}{\rm Q} \else \mbox{$\hspace{.2em}\vbar\hspace{-.31em}{\rm Q}$} \fi}
\newcommand{\Z}{\ifmmode {\rm Z}\hspace{-.28em}{\rm Z} \else ${\rm Z}\hspace{-.28em}{\rm Z}$ \fi}

\renewcommand{\bf}[1]{\mathbf{#1}}   

\renewcommand{\vec}[1]{\mathbf{#1}}     

\usepackage{xparse}
\DeclareDocumentCommand\cc{ g g }{%
	\mathbb{C} \IfNoValueF {#1}    {^{#1%
			\IfNoValueF {#2} { \times #2}%
	}}%
}

\DeclareDocumentCommand\rr{ g g }{%
	\mathbb{R} \IfNoValueF {#1}    {^{#1%
			\IfNoValueF {#2} { \times #2}%
	}}%
}

\DeclareDocumentCommand\set{ m g g}{%
	\{#1\}_{1\leq #2  	\IfNoValueF {#3} { \leq #3}}
}

\begin{document}

\IEEEoverridecommandlockouts

\title{Two-stage Beamformer Design via Deterministic Equivalents\\
	\thanks{This work has been supported in part by the Academy of Finland 6Genesis Flagship (grant no. 318927).}
}

\author{\IEEEauthorblockN{Hossein Asgharimoghaddam and Antti T\"olli}
	\IEEEauthorblockA{\textit{Centre for wireless communications (CWC),
		University of Oulu,	Oulu, Finland} \\
	\{hossein.asgharimoghaddam, antti.tolli\}@oulu.fi}
}

\maketitle

%

\maketitle
\vspace{-2cm}
\begin{abstract}
	Complexity reduction of optimal linear receiver is
	considered in a scenario where both the number of single antenna
	user equipments (UEs) $K$ and base station (BS) antennas $N$
	are large. Two-stage beamforming (TSB) greatly alleviates the
	high implementation complexity of large scale multiantenna
	receiver by concatenating a statistical outer beamformer (OBF)
	with an instantaneous inner beamformer (IBF) design. Using asymptotic large system analysis, we propose a novel TSB
	method that adjusts the dimensions of user specific OBF matrices
	based on the projection of the optimal minimum mean square
	error (MMSE) vectors into the beam domain. The beam
	domain is first divided into $S$ narrow sectors such that each
	sector contains $D$ DFT beams. Then, so called deterministic
	equivalents are computed for the amplitude-projection of the
	optimal MMSE vectors into each sector in asymptotic regime
	where $N$ , $K$ and $D$ grow large with a non-trivial ratio $N/K = C$
	and $N/D = S$. Given the approximations for the sector specific
	values, the structure and dimension of each UE specific OBF
	vector are optimized based on the statistical channel properties
	and the amount of overlap among users in angular domain. The
	numerical analysis shows that the attained SINR values closely
	follow the optimal MMSE receiver while the computational
	burden is greatly reduced.
\end{abstract}

\section{Introduction}
High spatial utilization is a promising approach to meet the significant spectral efficiency enhancements required for 5G cellular networks. In general, this is achieved by using a large number of antennas $N$ at the base stations (BSs) to serve a large number of user equipments (UEs) $K$ on the same frequency-time resources. 
However, such large dimensions of the channel matrices pose challenges on computational complexity and hardware costs. A promising solution to these problems lies in the concept of two-stage beamforming (TSB), which concatenates an outer-beamformer (analog/digital) with an inner beamformer/receiver.

Joint spatial division and multiplexing (JSDM) 
is introduced in~\cite{UserPartionAdhikary} for a downlink scenario wherein a statistical OBF matrix creates multiple virtual sectors. 
Exploiting the similarity among covariance matrices of co-located UEs, the authors in~\cite{UserPartionAdhikary} propose to group UEs based on their statistical properties. Then, the OBF matrix is designed based on the eigenvectors of group-specific covariance matrices.
 The performance of such a system depend on group-formation, and cross-sector interference management~\cite{XuJSDM-Grouping14,NamJSDM-Grouping14}. 
The authors in~\cite{AnttiGesbert16,AyshvarAntti18} study JSDM-based TSB in a downlink system to maximize the weighted sum-rate. It is observed that the reduced spatial dimensions results in significant inter-sector interference leakage as the number of UEs $K$ increases.
This issue is addressed in~\cite{AyshvarAntti18} by coordination of interference among sectors.
Similar performance  degradation appears in the equivalent uplink problem where the work in~\cite{TakahashiAntti19} mitigates the effects of inter-group interference using layered belief propagation detector.



In this paper, we consider uplink of a single-cell system wherein $K$ single-antenna UEs communicate with a BS equipped with $N$ antennas.
In this case, it is well-known that linear minimum mean square error (MMSE) receiver 
 attains the maximum signal-to-interference-plus-noise ratios (SINRs)~\cite{tse_viswanath_2005}. Motivated by this observation, a novel JSDMA-TSB method  is proposed 
that adjusts the dimensions of UE-specific OBF matrices
based on the projection of the MMSE vectors into the beam domain.
To this end, the angular-domain is divided into $S$ fixed narrow-sectors such that each sector contains $D<<N$ DFT beams.
  Then, so-called deterministic equivalents~\cite{RMT} are computed for the amplitude projection of MMSE vectors into each sector via asymptotic analysis in a regime, where $N$, $K$ and $D$ grow large with a non-trivial ratio $N/K=c$ and $N/D=S$.  The deterministic equivalents provide tight approximations for the considered metrics in finite-dimensional problems while those depend only on the statistical CSI~\cite{RMT}. The OBF matrix of each UE is obtained by concatenating the sectors whose AP-MMSES values are larger. As a result, the OBF matrices adopt the MMSE strategy, and adjust the direction and the number of sectors for each UE based on the level of multiple access interference while relying solely on  statistical CSI. The inner-receiver for a UE k is designed, as in the conventional TSB methods, based on  the resulting reduced dimensional channel matrix of size $D_k\times K$. 
  The numerical analysis shows that the attained per-UE rates closely follow the rate of optimal MMSE receiver. Also, it is observed that the dimension $D_k$ depends on the angular position of UE $k$, system load, UEs' angular spread, UEs' powers, and the desired bound on performance degradation.


\section{Problem Statement}
\label{sec:System model}
\subsection{System Model}
We consider uplink of a single-cell multi-user large-scale
MIMO system, where a single base station (BS) with $N$
antenna elements serves $K<N$ single-antenna user terminals (UE).
Under this convention and assuming narrow-band transmission, we define ${\bf h}_{k} \in \mathbb{C}^{N}$ as the channel between the BS and UE $k$. Then, the received signal of UE $k$ at BS can be expressed
as
\begin{equation}
\textstyle \ y_{k}  = {\bf w}_{k}\herm {\bf h}_{k} x_{k} +\sum_{i\setminus k}{\bf w}_{k}\herm{\bf h}_{i}s_{i} +{\bf w}_{k}\herm  {\bf n}\
\end{equation}
where the first term is the desired signal and the second
term represents intra-cell interference. 
The vector ${\bf w}_{k}\in \cc{N}$ denotes the  receiver  vector of UE $k$.
 The zero mean, unit variance data symbol intended to UE $k$ is denoted by  $x_{k}$, and is assumed to be independent across UEs.
  Zero-mean white  Gaussian noise at the receiver is denote by ${\bf n}\sim \mathcal {CN}(0,\sigma^{2}\eye{N})$.
  The MMSE receiver for a UE $k$ is given as $\mathbf{w}^{\star}_{k}=(\sum_{j\setminus k}{p_{j}}\mathbf{  h}_{j}\mathbf{  h}_{j}\herm + \sigma^2\mathbf{I}_{N})^{-1}\mathbf{  h}_{k}$ with $p_j$ being the transmit power of UE $j$. The superscript $()^{\star}$ indicates the optimality of the MMSE receiver.
  Nevertheless, the implementation of large scale MMSE receiver is not feasible with large antenna arrays due to computational complexity constraints.  
    Given 
    a limited angular spread\footnote{In a typical cellular configuration with a tower-mounted BS and no
  	significant local scattering, the propagation between the BS antennas and any
  	given UE is expected to be characterized by the local scattering around the
  	UE.  This results in UE's signal to arrive at BS from a limited angular spread.},
   it is possible to do receive-processing in a smaller dimensional space than $N$.   
  To this end, we first need to introduce a statistical model for the channel vectors.
  
  \subsection{Channel Model}
  \label{sec:Ch model}
  {The channel from BS to UE $k$ is modelled as ${\bf h}_{k} = {\bf \Theta}_{k}^{1/2}{\bf z}_{k}$ where 
  	${\bf z}_{k}\in  \mathbb {C}^{N}$ represents small-scale fading and has i.i.d, zero-mean, unit-variance complex entries.}
  The matrix ${\bf \Theta}_{k}\in  \mathbb {C}^{N\times N}$ accounts for the UE specific channel correlation at the BS. {The pathloss due to large scale fading is implicitly considered in the correlation matrix unless otherwise stated. In the latter case, pathloss values are explicitly declared by expressing the correlation matrix as $a^2_{k}{\bf \Theta}_{k}$  where $a^2_{k}$ accounts for pathloss from the BS to UE $k$. }

  \subsection{Beamformer design}
  \label{sec:Beamformer design}
   \begin{figure}[t]
  	\centering
  	\includegraphics[width=\linewidth]{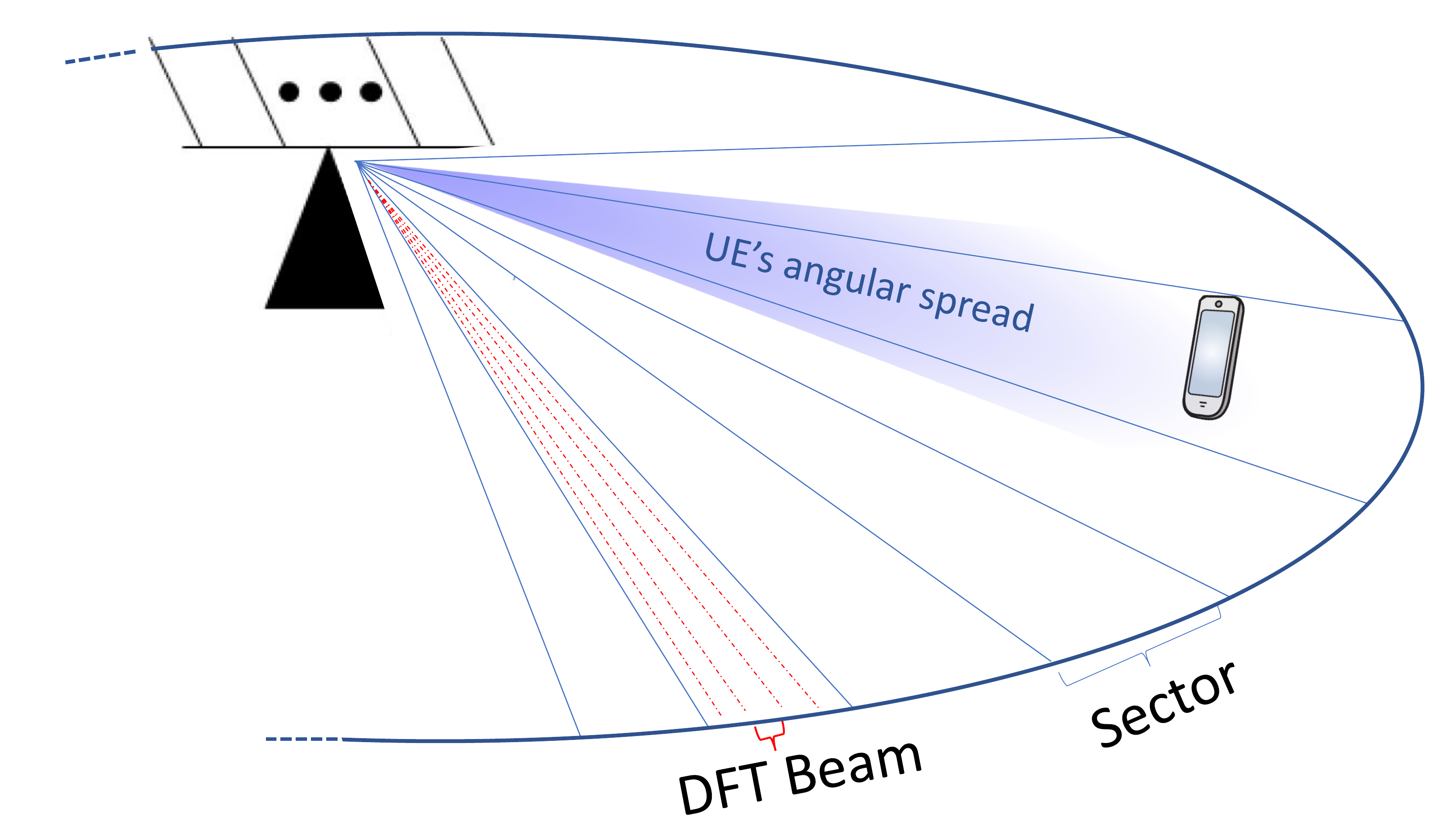}
  	\caption{Illustration of sectorized angular domain.}
  	\label{fig:sec}
  \end{figure}%
The beamforming vector of a UE $k$ is presented as $\vec{w}_k=\vec{B}_k\vec{v}_k$, where we'd like to design outer-beamfomer (OBF) $\vec{B}_k\in\cc{N}{D_k}$ based on statistical CSI in order to decrease the complexity of inner-receiver $\vec{v}_k \in\cc{D_k}$. Here, $\{D_k\}_{ \forall k}$ are UE dependent design parameters, which trade-off performance and complexity of obtaining the inner-receivers. 
 To do so, we divide the beam-domain into $S$ fixed narrow sectors $\set{\bf{S}_i}{i}{S}$ as shown in Fig.~\ref{fig:sec}.
Let   $\bf U$ be the $N\times N$ unitary matrix the columns of which are DFT vectors/beams $\set{\bf{u}_j}{j}{N}$.   Each sector contains $D$ DFT beams, i.e., $   \bf{S}_i=\{\bf{u}_j, j\in\{(i-1)\times D+1 ,...,(i-1)\times D+D\}\}$.
Since MMSE vector is the optimal receiver for the considered system model, we propose to form the OBFs based on projection of the optimal MMSE vectors into each sector. 
The projection of MMSE vector of UE $k$ into a sector $\vec{S}_i$ is given as $\vec{S}_i\herm \vec{w}^{\star}_k$, and thus, the normalized squared norm of this projection, denoted by $\omega_{k,i}$, is given as
\begin{equation}
\label{eq:amp project}
\omega_{k,i}=\frac{1}{N}\mathbf{  h}_{k} \herm \boldsymbol{\Sigma}_{k} \vec{S}_i \vec{S}_i \herm 
\boldsymbol{\Sigma}_{k}  \mathbf{  h}_{k}  
\end{equation}  
where   $\boldsymbol{\Sigma}_{k}=\big(\sum_{j \setminus k}{{p}_{j}}\mathbf{  h}_{j}\mathbf{  h}_{j}\herm  +\mathbf{I}_N\big)^{-1}$.    
Given approximations for $\omega_{k,i}$ values, the OBF for UE $k$ can be attained by selecting the sectors that have larger projection norm.  This ensures that the inner-receiver has enough information to yield SINR values close to the optimal MMSE ones.  Under this convention, the SINR of a UE is attained via inner-receiver by processing signals received within  a vector space of size $D_k<N$.   
%
 In deriving the approximations for $\omega_{k,i}$ values, we use results from random matrix theory that allows approximating functional  of the random matrices by deterministic quantities~\cite{RMT}. These quantities depend  only on the underlying statistical properties,  and yield precise approximations for practical problems of finite dimensions.  The result of this analysis is presented in the following section.
  
\section{Large system analysis}
\label{sec: Large sys}
    In deriving the large system analysis, the following assumptions (widely used in the literature) are made to properly define the growth rate of system dimensions. 
\begin{assumption}\label{as:0}
	As $N\to \infty$, $\!0 < \!  \frac{N}{K} < \infty$, and $0 < \!  \frac{N}{S} < \infty$.
\end{assumption}
\begin{assumption}
	\label{as:2}
	The spectral norm of ${\bf \Theta}_{k}$ is uniformly bounded as $N\to\infty$, i.e., $
	\!	\lim \sup_{{N \to \infty}}$ $ \!\!\!\max_{\forall k} \{\left\|{\bf \Theta}_{k}\right\|\}\!\! < \infty.$
\end{assumption}
In order to ensure that the total power in the system does not grow unbounded as the number of UEs grow large, we normalize UEs' powers by the number of antennas $N$. Also, without loss of generality, the Gaussian noise variance is assumed to be one.
Following  the same approach as in~\cite{wagner2012large}, deterministic equivalents for $\omega_{k,i}$ terms can be derived in terms of statistical CSI.  The results are summarized in the following theorem.
\begin{theorem}\label{th:1}
	Under  Assumptions \ref{as:0}-\ref{as:2}, the following holds almost surely 
	\begin{equation}
	 \omega_{k,i}-\bar{ \omega}_{k,i} \rightarrow 0
	\end{equation}
 where  the values of $\bar{ \omega}_{k,i}$ can be evaluated as
 \begin{equation}\label{eq:Th1 eq0}
  [\bar{\omega}_{1,i},...,\bar{\omega}_{K,i}]=(\mathbf{I}_K-\mathbf{L})^{-1} {\vec b}_{i},\,\forall i\in\{1,...,S\}
 \end{equation}  
 where	
\begin{equation}\label{eq:Th1 eq1}
\left[\mathbf{L}\right]_{i,j}= \frac{1}{N}  \frac{  {\tr}\left(\boldsymbol{\Theta}_{i} \mathbf{T}\boldsymbol{\Theta}_{j}\mathbf{T}\right) }{(1/\bar p_j+  \bar m_{j})^2},
\end{equation}
and
\begin{equation}\label{eq:Th1 eq2}
\begin{aligned}
{\vec b}_{i}=\!\!\left[\frac{1}{N} {\tr} \left(\boldsymbol{\Theta}_{1} \mathbf{T}\vec{S}_i \vec{S}_i \herm \mathbf{T}\right),\ldots,\frac{1}{N} {\rm Tr}(\boldsymbol{\Theta}_{K} \mathbf{T}\vec{S}_i \vec{S}_i \herm \mathbf{T})\right]
\end{aligned}
\end{equation}
with ${\bf{T}}$ given by
\begin{align}\label{eq:Th1 eq3}
{\bf{T}} = \left(\frac{1}{N}\sum\limits_{j\in \mathcal U}\frac{ \bar p_j{\bf{\Theta}}_{j}}{1+ \bar p_j\bar m_{j}} + {\bf I}_N\right)^{-1}\!\!\!,
\end{align}	
and $\bar m_{j},\forall j\in \mathcal{U}$ are given as the fixed-point solution of $\bar m_{j}=\frac{1}{N}\tr(\boldsymbol{\Theta}_{j}{\bf{T}}),\forall j\in \mathcal{U}$.
\end{theorem}
\begin{IEEEproof} The proof is given in Appendix~\ref{th: Proof of Th1}.\end{IEEEproof}
The results of the theorem yield approximations for $\omega_{k,i},k\in\mathcal{U},i\in\{1,...,S\}$ values in finite regime. The results are utilized in the following to propose algorithms for obtaining OBF matrices. 

\section{Algorithms for designing two-stage beamformers}
\label{sec:alg for TSB}
 Given approximations for $\omega_{k,i}$ values, the OBF for UE $k$ can be attained by selecting the sectors that have larger projection norm, i.e.,  $\bf{B}_k=\{\bf{S}_i\}_{i\in\mathcal{B}_k}$ where $\mathcal{B}_k$ holds the indices of selected sectors. The received signal for UE $k$ after applying OBF is given as
  \begin{equation}
   \ \bf{B}_k\herm\bf{y}  =\bf{B}_k\herm  {\bf H}\vec{x} +\bf{B}_k\herm  {\bf n}\
  \end{equation}
 where  ${\bf H}=[\vec{h}_1,...,\vec{h}_K]$, and $\vec{x}=[x_1,...,x_K]\tran$. 
The inner-receiver of UE $k$ applies a MMSE vector based on the $D_k\times K$ equivalent channel given by  $\bf{B}_k\herm  {\bf H}$. Since we have $D_k=|\mathcal{B}_k|\times D$, the complexity of inner-receiver is determined by the cardinality of set $\mathcal{B}_k$. For a given UE $k$, we propose to select the sectors whose $\bar \omega_{k,i}$ values are larger than $\delta \max(\bar \omega_{k,1},...,\bar \omega_{k,S})$. The parameter $0\leq \delta\leq 1$ trades off the complexity and performance. Larger $\delta$ values yield smaller $D_k$ values but also degrades the performance. 
 These steps are summarized in Algorithm~\ref{alg:1}.
 \vspace{0.2cm}
 \begin{algorithm} [H]
	\caption{Two-stage beamforming algorithm.}
	\label{alg:1}
	
	\begin{algorithmic}[1]
		\LOOP
		\IF{Any change in the UEs' statistics or during the initial stage}
		\STATE Obtain $\bar \omega_{k,i}$ values from~\eqref{eq:Th1 eq0}.
		\STATE Obtain OBFs as $\bf{B}_k=\{\bf{S}_i\}_{i\in\mathcal{B}_k},\forall k\in \mathcal{K}$ where $\mathcal{B}_k=\{j|\bar \omega_{k,j}\geq \delta \max(\bar \omega_{k,1},...,\bar \omega_{k,S})\}$
		\ENDIF
		\STATE Obtain inner-receivers as $\mathbf{v}_{k}=(\sum_{j\setminus k}{p_{j}}\bf{B}_k\herm\mathbf{  h}_{j}\mathbf{  h}_{j}\herm \bf{B}_k+ \sigma^2\mathbf{I}_{D_k})^{-1}\bf{B}_k\herm \mathbf{  h}_{k}$.
		\ENDLOOP
	\end{algorithmic}
\end{algorithm} 

Concerning the complexity analysis, we notice that the evaluation of inner-receiver $\mathbf{v}_{k}$ involves a matrix inversion of size $D_k\times D_k$ with a complexity in the order of $\mathcal{O}(D_k^3)$. Due to the limited angular spread of UEs' signals, $D_k$ values are expected to be much smaller than $N$.
Concerning the calculation of approximate $\bar \omega_{k,i}$ values, we notice that $\bar \omega_{k,i}$ values in Step 3 of the algorithm are updated only when there are sufficient changes in CSI statistics, which vary at a	much slower rate than the fading CSI. 
The computation of approximate $\bar \omega_{k,i}$ values requires matrix inversion in $(\mathbf{I}_K-\mathbf{L})^{-1}$, and evaluation of $\{\bar m_k\}$ values. The complexity of evaluating the former one is of order $\mathcal{O}(K^3)$. The latter one is evaluated via a fixed point iteration with complexity of $\mathcal{O}( N^3)$ per-iteration.

\section{Numerical Analysis}
\label{sec:Simulation Results}
Monte Carlo simulations are now used to validate the performance of the proposed solution.
 By assuming a diffuse 2-D field of isotropic scatterers around the receiver~\cite{JakesMicrowavebook}, the correlation matrix for an antenna element spacing of $\Delta$ is given by
\begin{equation}\label{eq:corr model}
\left[{\bf{\Theta}}_{k}\right]_{j,i}=\frac{a_{k}^2}{\varphi_{k}^{\max}-\varphi_{k}^{\min}}\int_{\varphi_{k}^{\min}}^{\varphi_{k}^{\max}} \! e^{i\frac{2\pi}{w}\Delta (j-i){\rm cos}(\varphi)} \, \mathrm{d}\varphi
\end{equation}
where waves arrive with an angular spread $\Delta \varphi$ from $\varphi_{\min}$ to $\varphi_{\max}$. The wavelength is denoted by $w$, and the antenna element spacing is fixed to half the wavelength $\Delta=1/2w$.  
The UEs are distributed over a circle of radius 300m between angular position $\frac{\pi}{6}$ to $\frac{5\pi}{6}$. The angular separation between UEs are the same and equal to $\frac{2\pi}{3}\frac{1}{K}$.   The  angular spread $\Delta \varphi$ is the same for all UEs and equal to $\pi/10$.
Thus, increasing the number of UEs results in an increase in overlap among UEs' signals angle-of-arrivals (AOAs). 
The number of antennas at BS is fixed to $N=225$, and the angular domain is divided into $S=45$ sectors.
The effect of pathloss and additive noise is captured in received signal to noise ratio (SNR) at an antenna element of BS. The SNR is denoted by $\rho$ in the following.

\begin{figure}[t]
	\centering
	{\includegraphics[clip, trim=0cm 7cm 0cm 0cm, width=\columnwidth]{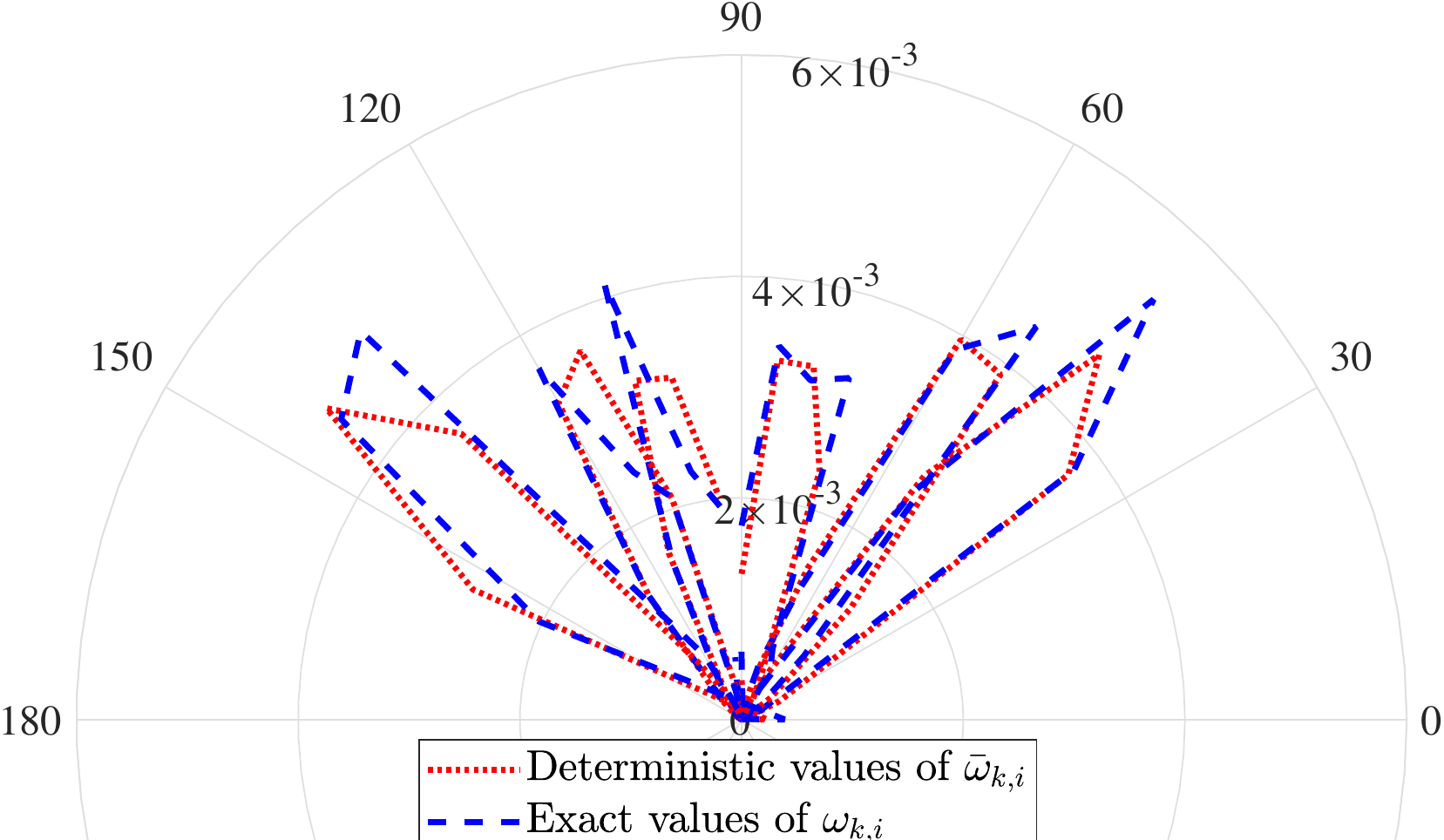}}
	\caption{The values of $\bar \omega_{k,i}$ and $\omega_{k,i}$ for 5 selected UEs, $N=225$, $K=135$, $S=45$.  }
	\label{fig:Accur}
\end{figure}

In order to validate the large system analysis, Fig.~\ref{fig:Accur} shows the exact values of $\omega_{k,i}$ and the deterministic equivalents $\bar \omega_{k,i}$ for 5 selected UEs, and a given random realization of small-scale fading.  The number of UEs $K$ is equal to $135$. It can be seen that the values of deterministic $\bar \omega_{k,i}$ closely follows the exact ones $\omega_{k,i}$. The OBF matrices in Algorithm~\ref{alg:1} are designed based on these accurate approximations. Thus, the spatial filtering is expected to reduce the dimensions of inner-receivers with minimal performance degradation.

 \begin{figure}[b]
	\centering
	\includegraphics[width=\linewidth]{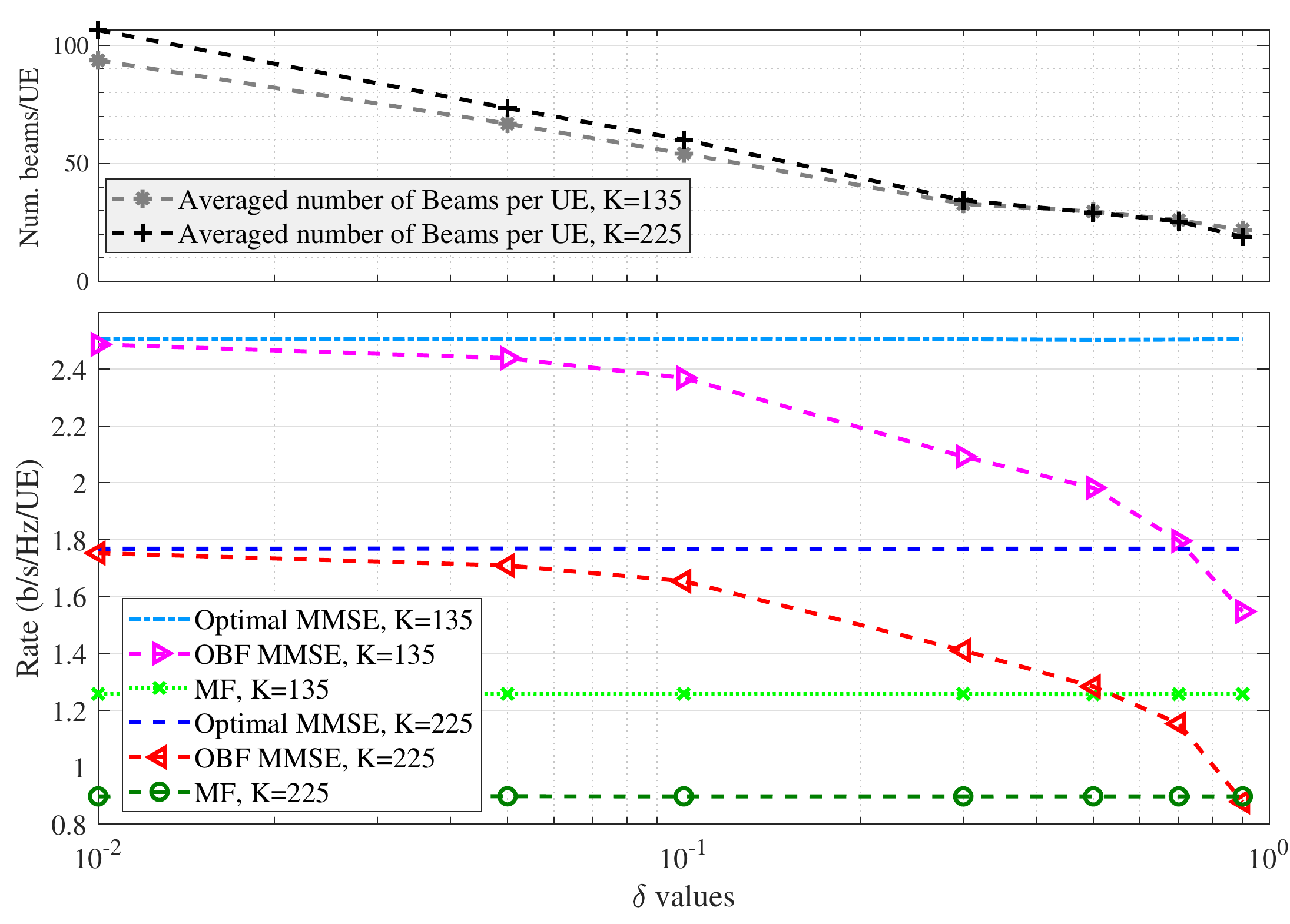}
	\caption{Rate and the number of beams per UE vs. $\delta$, $N=225$, $S=45$, $\rho=10$.}
	\label{fig:rate vs d1}
\end{figure}%
Fig.~\ref{fig:rate vs d1} illustrates the trade off between complexity and performance in Algorithm~\ref{alg:1}.
 The upper and lower plots in the figure show the averaged number of beams allocated per UE and attainable rates in b/s/Hz/UE, respectively, versus $\delta$ values.
 The results are presented for the cases with the number of UEs equal to 135 and 225.
 The attainable rate using Algorithm~\ref{alg:1} is titled as OBF-MMSE in the figure. Also, the rates of optimal MMSE receiver and matched filtering are presented as  benchmarks.
  As can be seen from the figure, the number of beams per UE decreases as $\delta$ value increases. The parameter $\delta$ adjusts the number of beams that are passed to the inner-receiver. A higher value of $\delta$ neglects more beams with small AP-MMSE values. Setting $\delta=0.1$ as an example, cuts off the sectors whose AP-MMSE values are less than one-tenth of the maximum value. It can be seen that at point $\delta=0.1$, the gap to the optimum rate is small. Also, the number of beams per UE is near $N/4$. Thus, at $\delta=0.1$ a proper trade off between the performance and complexity is achieved. 

\begin{figure}[t]
	\centering
	\includegraphics[width=\linewidth]{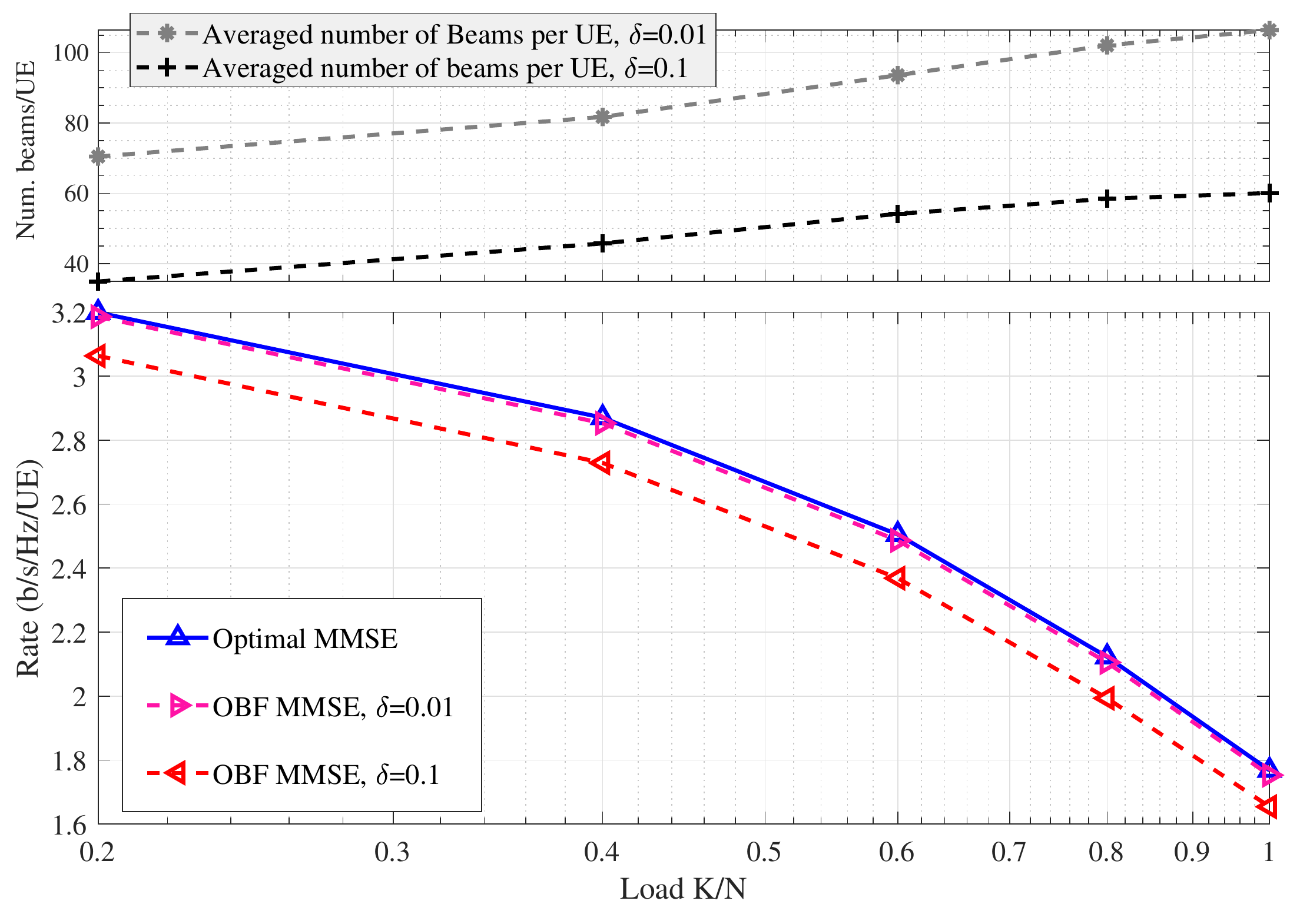}
	\caption{Rate and the number of beams per UE vs. load $K/N$, $N=225$, $S=45$, $\rho=10$.}
	\label{fig:rate vs KN}
\end{figure}
\begin{figure}[t]
 	\centering
 	\includegraphics[width=\linewidth]{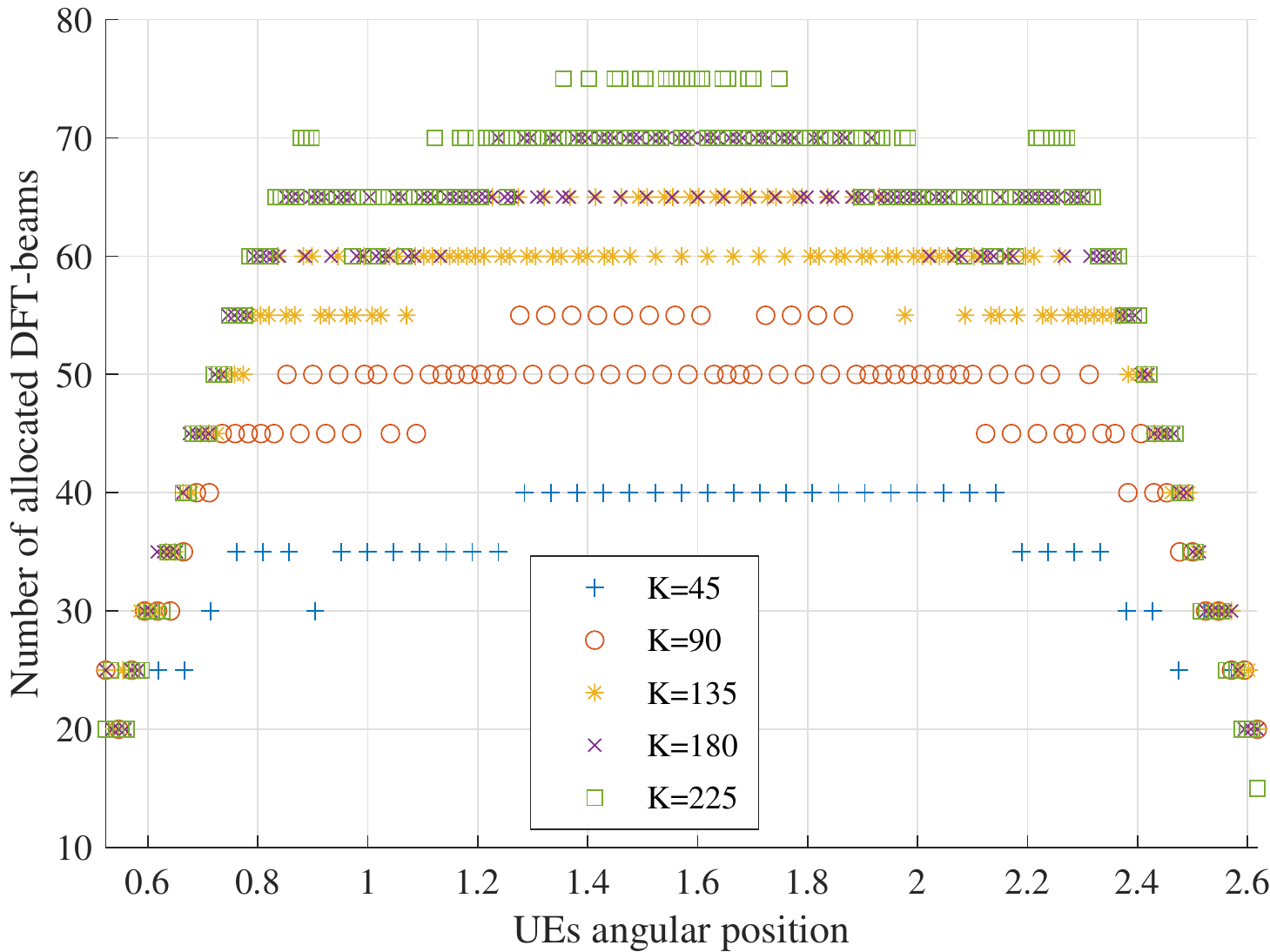}
 	\caption{The number of allocated beams vs. angular position of UEs, $\delta=0.1$, $N=225$, $S=45$, $\rho=10$.}
 	\label{fig:num beam vs ang}
 \end{figure}

In  Fig.~\ref{fig:rate vs KN}, the attainable rate in b/s/Hz/UE along with corresponding averaged number of beams per UE in  Algorithm~\ref{alg:1}  is plotted versus load $K/N$. The results are presented for the cases with $\delta$ equals to 0.01 and 0.1.
 Interestingly the gap to the optimal rate is almost fixed for a given value of $\delta$ over the whole range of load $K/N$. The larger values of $\delta$ yield a larger gap. It can be seen that the number of beams per UE increases as load of the system increases. This is due to the fact that larger load results in stronger multiple access interference. Thus, in order to keep the performance degradation within a given limit, a larger number of degrees of freedom is needed in the inner-receivers to mitigate the interference. In an alternative presentation, the number of beams allocated to each UE is plotted versus UEs' angular positions in Fig.~\ref{fig:num beam vs ang}. The value of parameter $\delta$ is fixed to $0.1$, and the results are plotted for various number of UEs. As mentioned earlier the higher load generally needs a larger number of beams to keep a certain performance degradation. The other observation is that the number of allocated beams is larger for  UEs residing in front of antenna array, while UEs in sides of the array needs smaller number of beams. This is due to the fact that the signal of UEs residing in sides of the array  are  less interfered. Also, the DFT beams become more dense in front of the array while the resolution of DFT beams decreases towards the sides of the array.

\section{Conclusions}
\label{sec:Conclusion} 
 Based on large system analysis, a novel TSB
method was proposed that adjusts the dimensions of UE-specific OBF matrices
based on the projection of the optimal MMSE vectors into the beam domain.  
This approach takes the multi-access interference into account when designing OBF, and thus, yields an optimal selection of sectors for a given UE.
This allowed us to study the optimal window-sizes $|\mathcal{B}_k|$  given a certain performance degradation. It was observed that the window-size in average increases as the load of the system grows large, i.e., as multiple-access interference increases.  Also, the numerical analysis showed that the UEs residing in the sides of the antenna array need smaller window-sizes, which is due to  lower interference and lower resolution of DFT beams in the sides of the array. It was shown that the attained SINR values based on the proposed approach closely
follow the optimal MMSE receiver while the computational burden of obtaining inner-beamformer is greatly reduced.

\appendices 
\section{Proof of Theorem~\ref{th:1}}
\label{th: Proof of Th1}
In the derivation of large system analysis, we use well-known lemmas including 
trace lemma~\cite[Lemma 2.6]{bai1998no},\cite[Theorem 3.4]{RMT} along with rank-1 perturbation lemma~\cite[Lemma 2.6]{silverstein1995empirical},\cite[Theorem 3.9]{RMT}. The former one shows asymptotic convergence of $\textbf{x}\herm \textbf{A} \textbf{x}-\frac{1}{N}\text{Tr}{\bf{A}}\rightarrow0$ when ${\bf x}\in \mathbb{C}^N$ has i.i.d entries with zero mean, variance of $\frac{1}{N}$ and independent of $\mathbf{A}$. The latter one states that addition of rank-1 matrix ${\bf {xx}}\herm$ to the random Gram matrix ${\bf{X}} {\bf X}\herm$ does not affect trace $\frac{1}{N}\text{Tr}({\bf X} {\bf X}\herm+{\bf I}_N)$ term in the large dimensional limit. The formal presentation of these lemmas are given in~\cite{bai1998no,silverstein1995empirical}. 

Starting from amplitude projection in~\eqref{eq:amp project}, we apply trace lemma, along with rank-1 perturbation lemma to get
\begin{equation}\label{eq:proof Th1 eq 1}
\begin{aligned}
\omega_{k,i}-\frac{1}{N}\tr( {\bf \Theta}_{k}\boldsymbol{\Sigma} \vec{S}_i \vec{S}_i \herm \boldsymbol{\Sigma}  )\xrightarrow{N\rightarrow \infty} 0 \\
\end{aligned}
\end{equation}
almost surely, where  $\boldsymbol{\Sigma}=\big(\sum_{j }{{p}_{j}}\mathbf{  h}_{j}\mathbf{  h}_{j}\herm  +\mathbf{I}_N\big)^{-1}$.
From matrix identities~\cite{Horn-Johnson-90}, we know that $ 
{\partial \mathbf{Y}^{-1}}/{\partial x}=- \mathbf{Y}^{-1} ( {\partial \mathbf{Y}}/{\partial x} ) \mathbf{Y}^{-1}
$  with $\bf Y$ being a matrix depending on variable $x$. Thus, the above trace term can be written equivalently as
\begin{equation}\label{eq:proof Th1 eq 2}
\frac{1}{N}\tr( {\bf \Theta}_{k}\boldsymbol{\Sigma} \vec{S}_i \vec{S}_i \herm \boldsymbol{\Sigma}  )=\frac{\partial}{\partial x} m_{k,i}(z,x)|_{x=0,z=-1}
\end{equation}
where 
\begin{equation}
m_{k,i}(z,x)=\frac{1}{N} \tr\big(\boldsymbol{\Theta}_{k}
\big(\sum_{j }{{p}_{j}}\mathbf{  h}_{j}\mathbf{  h}_{j}\herm  -z\mathbf{I}_N-x \vec{S}_i \vec{S}_i \herm  \big)^{-1} \big).
\end{equation}
The term $m_{k,i}(z,x)$ is the Stieltjes transforms of a measure. It is shown in~\cite[Theorem 1]{wagner2012large}, where
 under Assumption~\ref{as:0}-\ref{as:2}, and for $z \in \mathbb{C} \backslash \mathbb{R}^+$, $x\in \mathbb{R}^-$, these Stieltjes transforms have deterministic equivalents such that 
\begin{equation}\label{eq:proof Th1 eq 3}
m_{k,i }(z,x)-\bar{m}_{k,i }(z,x)\xrightarrow{N\rightarrow \infty} 0
\end{equation}
almost surely,
where the deterministic equivalents $\bar{m}_{k,i }(z,x)$ are given as the  solutions of the following fixed-point iterations
\begin{equation}\label{Proof equivalent ST form}
\bar{m}_{k,i}(z,x)=\frac{1}{N} \!{\tr }\big(\boldsymbol{\Theta}_k \bf{T}_i(z,x)\big), \forall k\in\mathcal{U}, i\in\{1,...,S\}
\end{equation}
where
\begin{equation}
\bf{T}_i(z,x)=\biggr(\frac{1}{N}\sum\limits_{j=1}^K \frac{p_j \boldsymbol{\Theta}_j}{1+p_j\bar{m}_{j,i}(z,x)}-x\vec{S}_i \vec{S}_i \herm-z\mathbf{I}_N\biggr)^{-1}.
\end{equation}
As the result, from~\eqref{eq:proof Th1 eq 1},~\eqref{eq:proof Th1 eq 2}, and~\eqref{eq:proof Th1 eq 3}, we get
\begin{equation}
\omega_{k,i}- \bar m^{\prime}_{k,i}\xrightarrow{N\rightarrow \infty} 0
\end{equation}
where $\bar m^{\prime}_{k,i}= \bar m^{\prime}_{k,i}(z,x)|_{x=0,z=-1}$ with $\bar m^{\prime}_{k,i}(z,x)\triangleq\frac{\partial}{\partial x} \bar m_{k,i}(z,x)$. The values of $\bar m^{\prime}_{k,i}$ can be evaluated by taking derivative of $\bar{m}_{k,i}(z,x)$ in~\eqref{Proof equivalent ST form}, and evaluating the derivative at point $({x=0,z=-1})$. In doing so, we get 
\begin{equation}
\bar{m}^{\prime}_{k,i}=\frac{1}{N} \!{\tr }\big(\boldsymbol{\Theta}_k \bf{T}^{\prime}_i\big), \forall k\in\mathcal{U}, i\in\{1,...,S\}
\end{equation}
where $\bf{T}^{\prime}_i=\bf{T}^{\prime}_i(z,x)|_{x=0,z=-1}$, or equivalently
\begin{equation}\label{eq:Tbkl prime}
\mathbf{T}_{i}^{\prime} = \mathbf{T}\bigg(\frac{1}{N} \sum_{j \in \mathcal{U}} \frac{{p}_j^2 \boldsymbol{\Theta}_{j}\bar{m}_{j,i}^{\prime}}{(1+{ p}_j\bar{ m}_{j})^2} +\vec{S}_i \vec{S}_i \herm \bigg) \mathbf{T}
\end{equation}
where $\mathbf{T}=\mathbf{T}_{i}(-1,0)$, and $\bar{ m}_{j}=\bar{ m}_{j,i}(-1,0)$ . Since $\bar{m}_{k,i}'=\frac{1}{N} \text{Tr} (\boldsymbol{\Theta}_{k} \mathbf{T}_{i}^{\prime})$ with $\mathbf{T}_{i}^{\prime}$ given by~\eqref{eq:Tbkl prime}, we get a system of equation to evaluate $\bar{m}_{k,i}'$ as $  [\bar{m}_{1,i}^{\prime},...,\bar{m}_{K,i}^{\prime}]$ $=(\mathbf{I}_K-\mathbf{L})^{-1} {\vec b}_{i}$ with ${\vec b}_{i}$ and $\mathbf{L}$ defined as in~\eqref{eq:Th1 eq2} and \eqref{eq:Th1 eq3}, respectively, which completes the proof of the theorem.

\bibliographystyle{IEEEtran}
%
%
\bibliography{Jourbib}
%
%
%

\end{document}